\documentclass{elsart}
\usepackage{natbib,epsfig}
\begin{document}
\runauthor{Lessard, Buckley, Connaughton and Le Bohec}
\begin{frontmatter}
\title{A New Analysis Method for Reconstructing the Arrival Direction
of TeV Gamma-rays Using a Single Imaging Atmospheric Cherenkov Telescope}
\author[Purdue]{R.W. Lessard}
\author[Washington]{J.H. Buckley}
\author[NASA]{V. Connaughton}
\author[ISU]{S. Le Bohec}

\address[Purdue]{Department of Physics, Purdue University, 
		West Lafayette, IN, 47907, USA}
\address[Washington]{Department of Physics, Washington University, 
		St. Louis, MO, 63130, USA}
\address[NASA]{NASA, Marshall Space Flight Center, Huntsville, AL, 35812, USA}
\address[ISU]{Department of Physics and Astronomy, Iowa State University,\\
		Ames, IA, 50011, USA}
\begin{abstract}
We present a method of atmospheric Cherenkov imaging which
reconstructs the unique arrival direction of TeV gamma rays using a
single telescope. The method is derived empirically and utilizes
several features of gamma-ray induced air showers which determine, to
a precision of $0.12^\circ$, the arrival direction of photons, on an
event-by-event basis. Data from the Whipple Observatory's 10 m
gamma-ray telescope is utilized to test selection methods based on
source location.  The results compare these selection methods with
traditional techniques and three different camera fields of view. The
method will be discussed in the context of a search for a gamma-ray
signal from a point source located anywhere within the field of view and
from regions of extended emission.
\end{abstract}
\end{frontmatter}

\section{Introduction}
The Whipple Collaboration operates a 10 m optical reflector for
gamma-ray astronomy at the Fred Lawrence Whipple Observatory on
Mt. Hopkins (elevation 2320 m) in southern Arizona. The reflector was
originally constructed in 1968 \cite{weekes72} and numerous
modifications have been made to improve the sensitivity and
performance of the system. A camera consisting of photomultiplier
tubes (PMTs) mounted in the focal plane of the reflector, detects the
Cherenkov radiation produced by gamma-ray and cosmic-ray air showers
from which an image of the Cherenkov light can be reconstructed.  The
camera is triggered when any two PMT signals are above a threshold
within a short time coincidence.

\begin{figure}
\epsfig{file=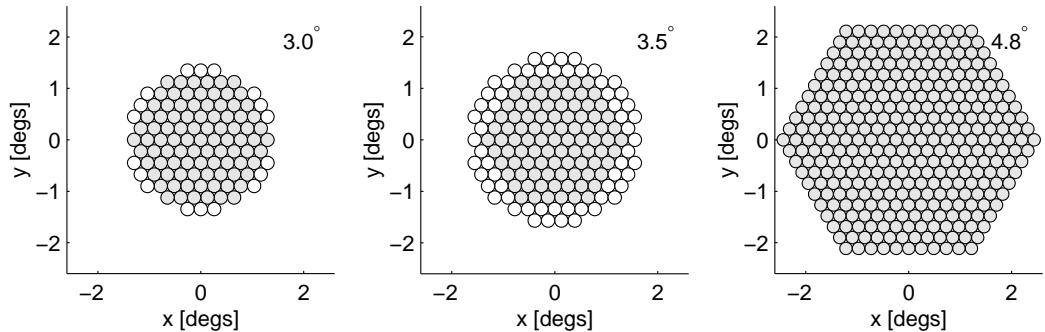,width=\linewidth}
\caption{The PMT camera utilized by the Whipple Observatory's 10 m
gamma-ray telescope has undergone several upgrades over the past
decade. Each camera is comprised of PMTs of the same model and size
but with increasing FOV. The gray pixels depict the PMTs included in
the trigger.}
\label{figure:camera}
\end{figure}

In the past decade the camera has undergone significant expansion from
a field of view (FOV) of $3.0^\circ$ in 1989 to a FOV of $4.8^\circ$
in 1999. From 1989 to 1996 the camera consisted of 109 PMTs, each
viewing a circular field of $0.259^\circ$ diameter, yielding a total
FOV of $3.0^\circ$. The trigger condition required any two of the
inner 91 PMT signals to be above a threshold within a 15 ns time
coincidence. During the spring of 1997 the camera was expanded to 151
PMTs yielding a total FOV of $3.5^\circ$. The trigger condition
required any two of the inner 91 PMT signals ($3.0^\circ$ triggering
FOV) to be above a threshold with a coincidence resolving time of 15
ns.  From 1997 to 1999 the camera was further expanded to 331 PMTs
yielding a total FOV of $4.8^\circ$. The trigger was expanded to
include all 331 PMT signals and the time coincidence was shortened to
10 ns due to the introduction of constant fraction discriminators. The
layout of each camera is depicted in Figure~\ref{figure:camera}. A
full description of the reflector, which has not changed since its
original construction, can be found in \cite{cawley90}.

Primary cosmic rays and gamma rays entering the atmosphere initiate
showers of secondary particles which propagate down towards the
ground.  The trajectory of the shower continues along the path of the
primary particle.  If the optical reflector lies within the 300 m
diameter Cherenkov light pool, it forms an image in the PMT camera.
The appearance of this image depends upon a number of factors. The
nature and energy of the incident particle, the arrival direction and
the point of impact of the particle trajectory on the ground, all
determine the initial shape and orientation of the image. This image
is modified by the point spread function of the telescope, the
addition of instrumental noise in the PMTs and subsequent electronics,
the presence of bright stellar images in certain PMTs and by spurious
signals from charged cosmic rays physically passing through the
tubes. Monte Carlo studies have shown that gamma-ray induced showers
give rise to more compact images than background hadronic showers and
are preferentially oriented towards the source position in the image
plane \cite{hillas85}. By making use of these differences, a gamma-ray
signal can be extracted from the large background of hadronic showers
and a gamma-ray map over the FOV can be obtained. The method of
extracting a gamma-ray signal from the hadronic background can be
found in \cite{fegan94}. The technique of obtaining a gamma-ray map of
the FOV, by reconstructing the unique arrival direction of very high
energy photons, is the topic of this paper.

Data from the Whipple Observatory's 10 m high energy gamma-ray
telescope, will be used to demonstrate methods of the reconstruction
of the arrival direction of gamma-ray induced showers.  These methods
may be applied a) to a search for point sources located anywhere
within the camera's FOV, for example searches for counterparts to
EGRET unidentified sources, gamma-ray bursts, described in
\cite{connaughton98}, or sky surveys and b) analysis
of extended sources of TeV gamma-rays such as supernova remnants,
described in \cite{buckley98} and Galactic plane emission described in
\cite{lebohec2000}.

\section{Shower Image Processing and Characterization}
Prior to analysis of the recorded images, two calibration operations
must be performed: the subtraction of the pedestal analog-digital
conversion (ADC) values and the normalization of the PMT gains, a
process known as flat-fielding.

The pedestal of an ADC is the finite value which it outputs for zero
input. This is usually set at 20 digital counts so that small negative
fluctuations on the signal line, due to night sky noise variations,
will not generate negative values in the ADC. The pedestal for each
PMT is determined by artificially triggering the camera, thereby
capturing ADC values in the absence of genuine input signals. The PMT
pedestal and pedestal variance are calculated from the mean and
variance of the pulse-height spectrum (PHS) generated from these
injected events.

The relative PMT gains are determined by recording a thousand images
using a fast Optitron Nitrogen Arc Lamp illuminating the focal plane
through a diffuser. These nitrogen pulser images are used to determine
the relative gains by comparing the relative mean signals seen by each
PMT.

\begin{figure}
\epsfig{file=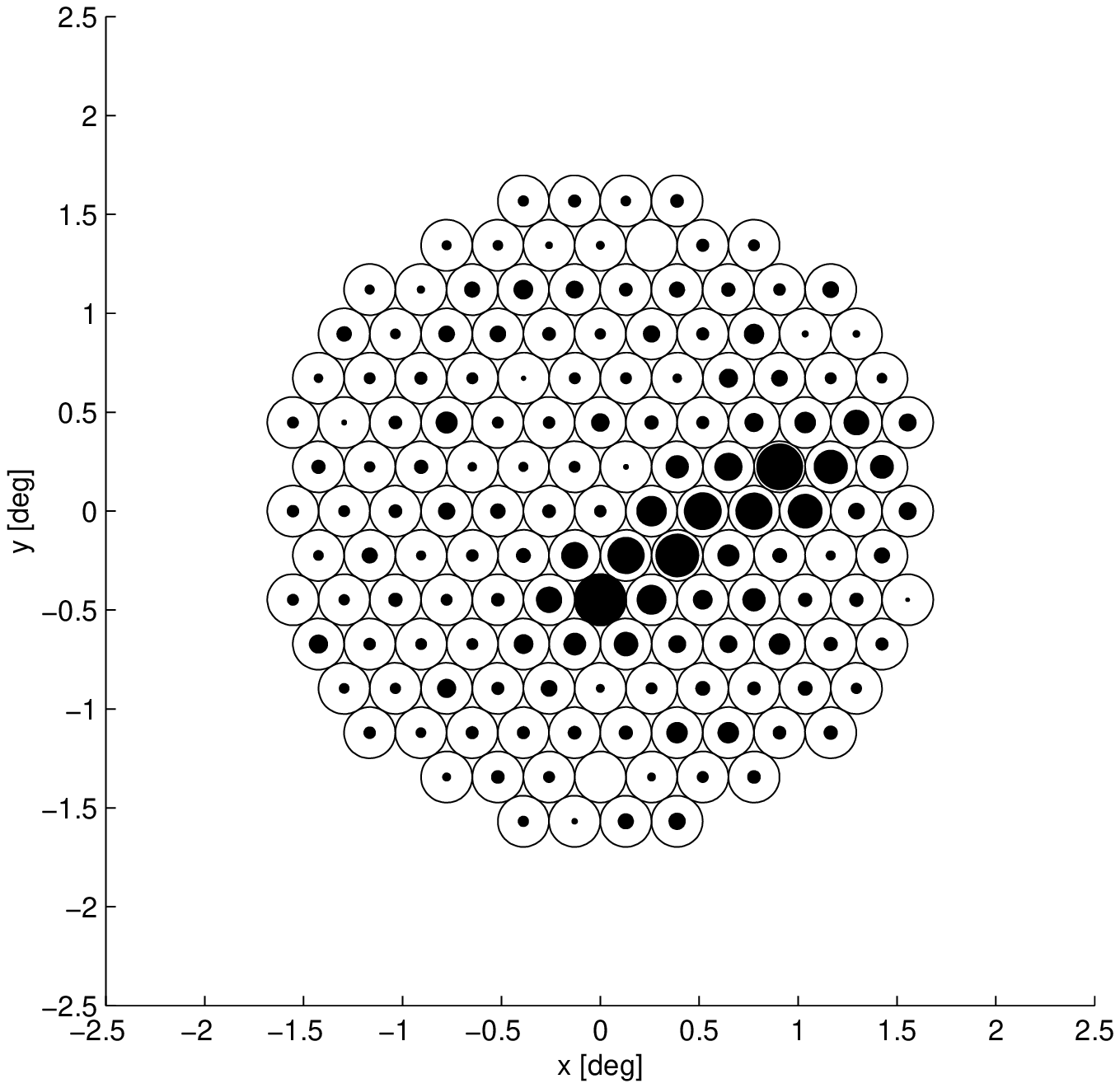,width=0.5\linewidth}
\epsfig{file=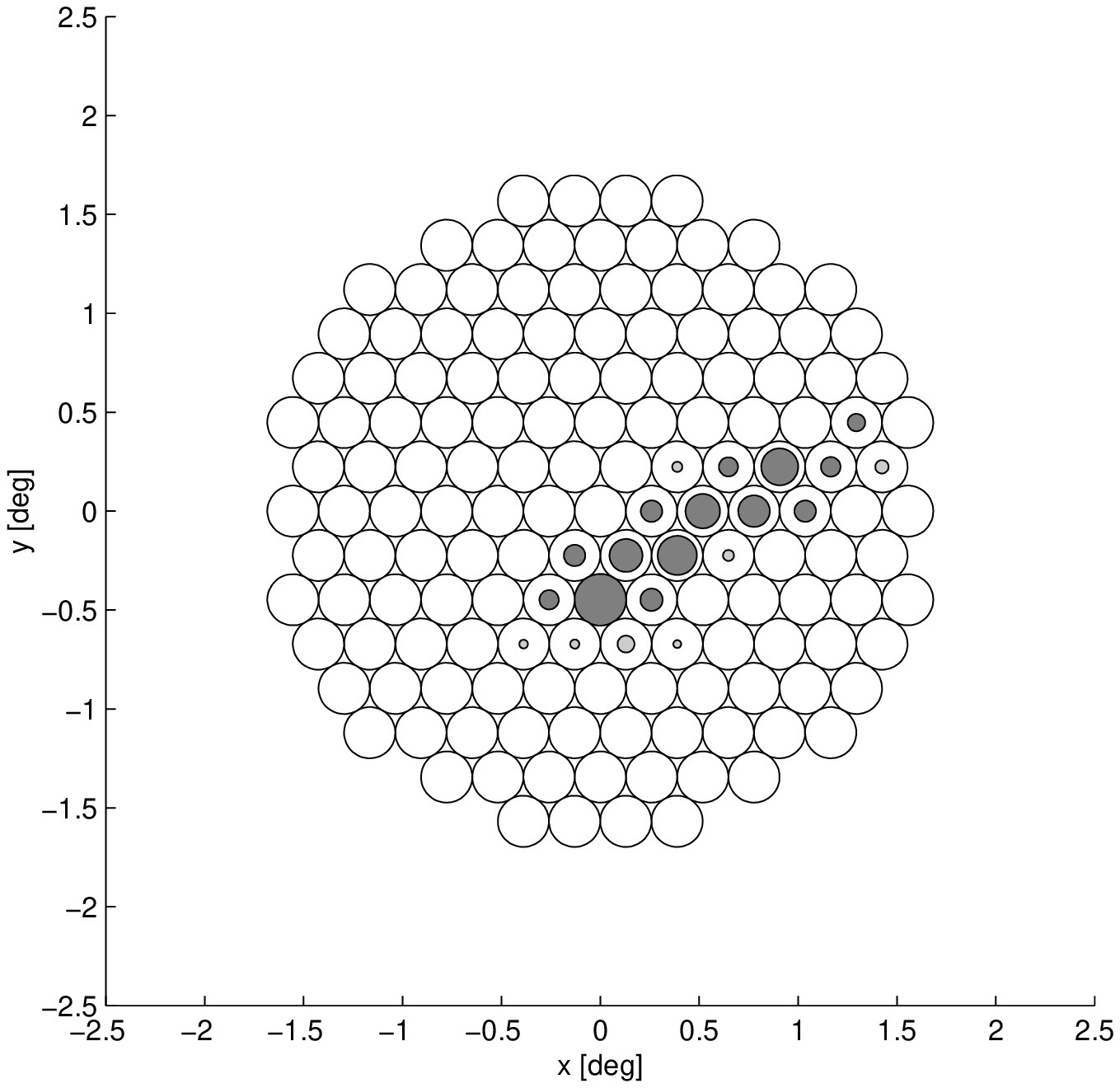,width=0.5\linewidth}
\caption{(Left) Example of an image prior to the processing procedure
given in the text. Diameter of each filled circle is proportional to
the ADC signal for that PMT. (Right) The same image after pedestal
subtraction, application of picture (dark gray filled circles) and
boundary (light gray filled circles) thresholds and gain
normalization.}
\label{figure:imclean}
\end{figure}

Fluctuations in the image usually arise from electronic noise and
night-sky background variations. To reject these distortions a PMT is
considered to be part of the image if it either has a signal above a
certain threshold or is beside such a PMT and has a signal above a
lower threshold. These two thresholds are defined as the picture and
boundary thresholds, respectively. The picture threshold is the
multiple of the root mean square (RMS) pedestal deviation which a PMTs
signal must exceed to be considered part of the picture. The boundary
threshold is the multiple of the RMS pedestal deviation which PMTs
adjacent to the picture must exceed to be part of the boundary. The
picture and boundary PMTs together make up the image; all others are
zeroed. This image cleaning procedure is depicted in
Figure~\ref{figure:imclean}. These thresholds were optimized using
data taken on the Crab Nebula yielding picture threshold: 4.25, and
boundary threshold: 2.25.

For the results reported in this paper, the data are obtained in an
ON/OFF mode where the source position is tracked for 28 minutes (ON),
followed by a 28 minute observation of a background region (OFF)
covering the same path in elevation and azimuth. In an ON/OFF
observation mode, differences in sky brightness between ON and OFF sky
regions could introduce biases. These biases can severely affect the
selection of pixels accredited to the boundary region and hence
distort the Cherenkov image. For example, if the ON region is brighter
than the OFF region, bias may arise as follows. Consider tubes which
have a combination of small amounts of genuine signal coupled with
some noise (e.g. boundary tubes). If the noise level is low, the
boundary threshold is low and most of these tubes will pass the
threshold test and be included as part of the image. However, if the
noise level is high then the boundary threshold will also be high and
the probability increases that a negative noise fluctuation will
cancel the genuine signal component resulting in the tube being set to
zero during image cleaning. Thus, the degree to which boundary tubes
are set to zero depend on the noise level in the tubes. As a
consequence a bright sky region will result in more boundary tubes
being set to zero, making the image appear narrower and more gamma-ray
like.

A software technique was developed to correct for the biases
introduced by the differing sky brightness \cite{cawley93}. This
technique, known as software padding works by adding software noise
into the events for the darker sky region. Let $P_{\rm on}(P_{\rm
off}), \sigma_{\rm on}(\sigma_{\rm off})$ be the ON(OFF) pedestal and
pedestal deviation values for a particular pixel and $C_{\rm
on}(C_{\rm off}), \sigma_{C_{\rm on}}(\sigma_{C_{\rm off}})$ be the 
component due to the Cherenkov signal and corresponding fluctuation. 
The total ON signal is
\begin{equation}
{\rm ON} = P_{\rm on} + \sigma_{\rm on} Gauss(0:1) + C_{\rm on} + 
\sigma_{C_{\rm on}} Gauss(0:1),
\end{equation}
where $Gauss(0:1)$ is a random number drawn from a Gaussian
distribution of zero mean and unit variance. The noise component due
to the night sky light in the ON region is:
\begin{equation}
N_{\rm on} = \sigma_{\rm on} Gauss(0:1).
\end{equation}

Similarly, for the OFF region:
\begin{equation}
N_{\rm off} = \sigma_{\rm off} Gauss(0:1).
\end{equation}

Suppose we are working with pairs where $N_{\rm on}$ is larger than
$N_{\rm off}$ then we wish to add noise, $N_{\rm add}$, in the OFF
events such that
\begin{equation}
N_{\rm on}^2 = N_{\rm off}^2 + N_{\rm add}^2,
\end{equation}
or
\begin{equation}
N_{\rm add} = \sqrt{N_{\rm on}^2 - N_{\rm off}^2}.
\end{equation}

The total OFF signal is then 
\begin{eqnarray}
{\rm OFF} = P_{\rm off} + \sigma_{\rm off} Gauss(0:1) + 
N_{\rm add} Gauss(0:1) + \nonumber \\
      C_{\rm off} + \sigma_{C_{\rm off}} Gauss(0:1).
\end{eqnarray}

When the OFF region is brighter than the ON region then $N_{\rm add}$
is added to the ON pixels. This software padding procedure is very
efficient in removing most of the biases induced by the sky brightness
with only a modest reduction in sensitivity. Given the potential for
large systematic errors without software padding, the modest reduction
in sensitivity is a small price to pay.

\begin{figure}
\epsfig{file=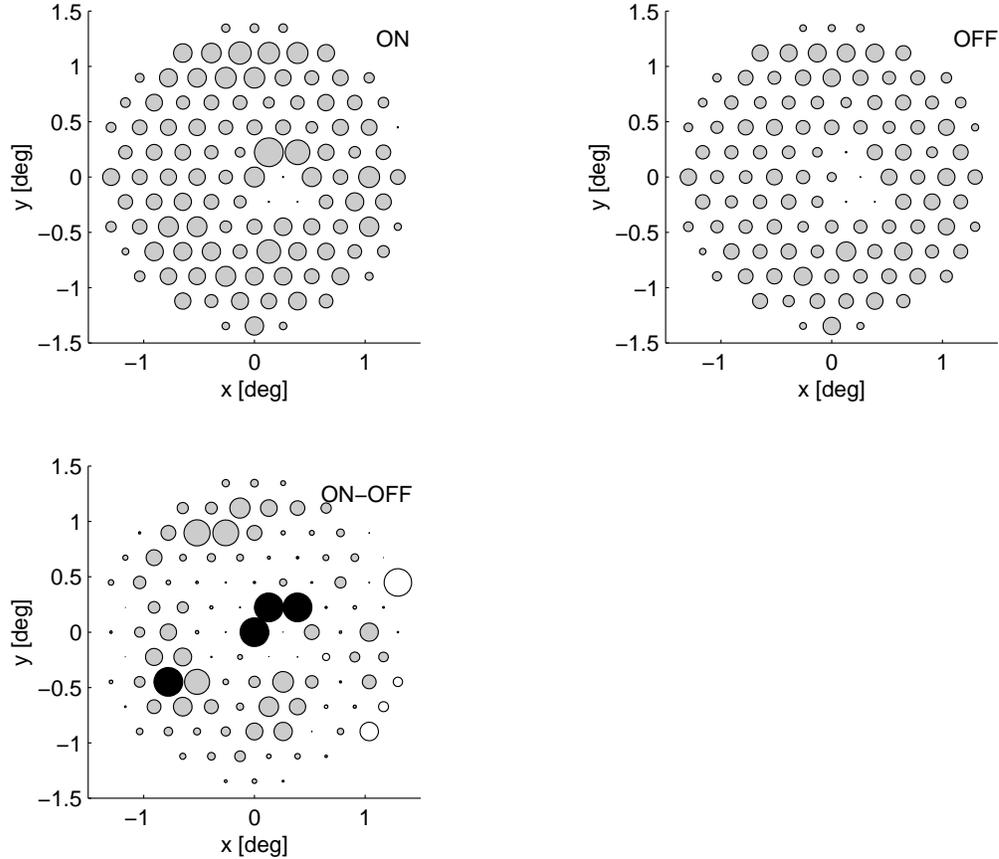,width=\linewidth}
\caption{Comparison of the sky brightness for the ON and OFF source
regions for supernova remnant G78.2+2.1 (a shell-type remnant located
in a bright region of the galactic plane). In the top panels, the size
of the circle is proportional to the RMS pedestal deviation for each
PMT, scaled to a maximum value of 9.2 d.c. The bottom panel depicts
the difference between the ON and OFF source observation, scaled to a
maximum difference of 2.0 d.c. Positive differences are shown as
filled circles, while open circles denote negative differences.  The
black circles show PMTs with a greater than 2.0 d.c. difference, due
to the presence of bright stars in the ON source region.}
\label{figure:gcygni_pedvar}
\end{figure}

An example of sky brightness is given in
Figure~\ref{figure:gcygni_pedvar} for the ON and OFF source regions of
the supernova remnant G78.2+2.1 (see \cite{buckley98} for the results
of TeV gamma-ray observations).  For this object, located in a bright
region of the galactic plane, the difference in sky brightness is
quite large thus the application of software padding is essential. The
results of the analysis of this extended object, with and without the
application of software padding, will be presented later in this
paper.

\begin{figure}
\epsfig{file=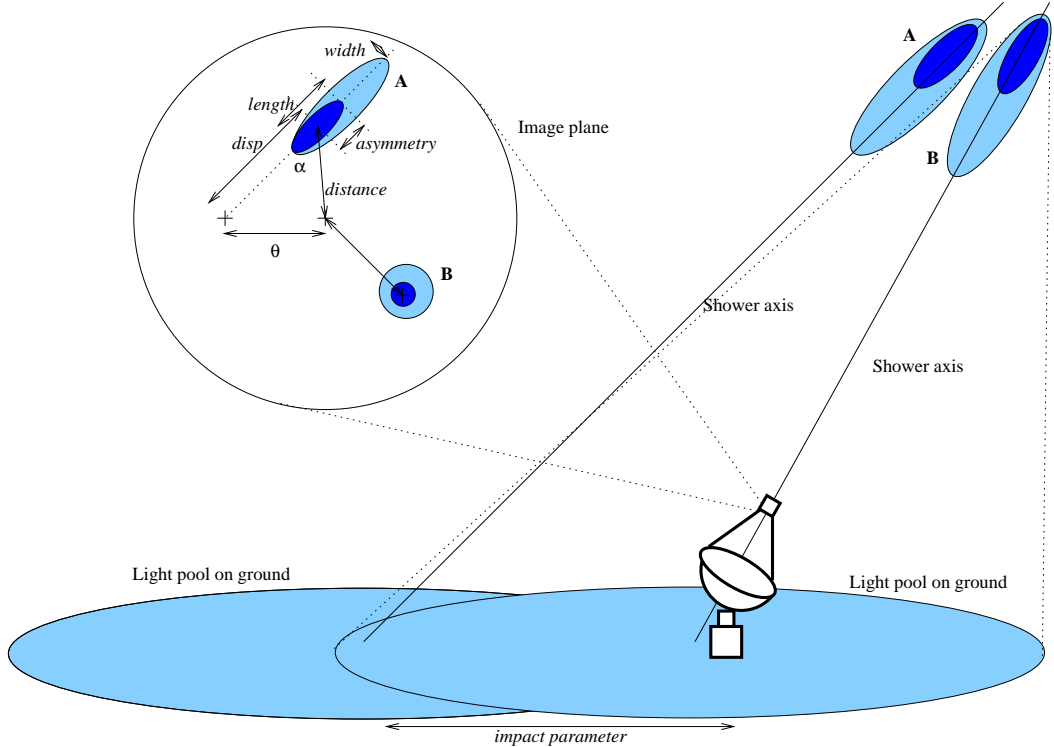,width=\linewidth}
\caption{Depiction of the light produced by air showers. The image
plane shows the definition of the Hillas parameters used to
characterize each image.}
\label{figure:hillasparameters}
\end{figure}

We characterize each Cherenkov image using a moment analysis
\cite{reynolds93}. The roughly elliptical shape of the image is
described by the {\em length} and {\em width} parameters. Its location
and orientation within the FOV are given by the {\em distance} and
$\alpha$ parameters, respectively. The {\em asymmetry} parameter,
defined as the third moment of the light distribution, describes the
skew of the image along its major axis. We also determine the two
highest signals recorded by the PMTs ({\em max1, max2}) and the amount
of light in the image ({\em size}). These parameters are defined in
Table~\ref{table:hillasparameters} and are depicted in
Figure~\ref{figure:hillasparameters}.

\begin{table}
\caption{Definition of image parameters, used to
characterize the image shape and orientation in the FOV (see
Figure~\protect\ref{figure:hillasparameters}).\label{table:hillasparameters}}
\vspace{0.5cm}
\begin{tabular}{ll}\hline
Parameter       & Definition                                                \\ 
\hline
{\em max1}:     & largest signal recorded by the PMTs.                      \\
{\em max2}:     & second largest signal recorded by the PMTs.               \\
{\em size}:     & sum of all signals recorded.                              \\
{\em centroid}: & weighted center of the light distribution ($x_c,y_c$).\\
{\em width}:    & the RMS spread of light along the minor axis of the image;\\
                & a measure of the lateral development of the shower.       \\
{\em length}:   & the RMS spread of light along the major axis of the image;\\
                & a measure of the vertical development of the shower.      \\
{\em distance}: & the distance from the centroid of the image to the center \\ 
                & of the FOV.                                               \\
$\alpha$:       & the angle between the major axis of the image and a line  \\
                & joining the centroid of the image to the center of the    \\
                & FOV.                                                      \\
{\em asymmetry}:& the skewness of the light distribution relative to the    \\
                & image centroid.                                           \\ 
{\em disp}:     & the angular distance from the image centroid to the assumed\\
                & arrival direction of the shower in the image plane.       \\
$\theta$:       & the angular distance from the arrival direction of the    \\
                & shower in the image plane and the center of the FOV.      \\
\hline
\end{tabular}
\end{table}

\section{Analysis Techniques}
Gamma-ray events give rise to shower images which are preferentially
oriented towards the source position in the image plane. These images
are narrow and compact in shape, elongating as the impact parameter
increases. They generally have a cometary shape with their light
distribution skewed towards their source location in the image
plane. Hadronic events give rise to images that are, on average,
broader (due to the emission angles of pions in nucleon collisions
spreading the shower), and longer (since the nucleon component of the
shower penetrates deeper into the atmosphere) and are randomly
oriented within the FOV. Utilizing these differences, a gamma-ray
signal can be extracted from the large background of hadronic showers.

\subsection{Traditional Methods}
The standard gamma-ray selection method utilized by the Whipple
Collaboration is the Supercuts criteria (see
Table~\ref{table:supercuts}; cf., \cite{reynolds93};
\cite{catanese95}; \cite{lebohec2000}). These criteria were optimized
on contemporaneous Crab Nebula data giving the best sensitivity to
point sources positioned at the center of the FOV. In an effort to
remove background events triggered by single muons and night sky
fluctuations, Supercuts incorporates pre-selection cuts on the {\em
size} and on {\em max1} and {\em max2}. The cuts on {\em width} and
{\em length} select the more compact gamma-ray images and the cut on
{\em distance} selects images for which the pointing angle $\alpha$ is
well defined. The changes to the upper bounds of the {\em length} and
{\em distance} cuts for the various cameras reflects the increasing
FOV which results in less image truncation thereby allowing an
accurate reconstruction of more distant images. A final cut on
$\alpha$ selects images which are aligned towards the source position,
assumed to be at the center of the FOV.  A gamma-ray signal is
detected as a statistically significant excess of events, which pass
the above criteria, between ON source and OFF source observations. In
the case when the source is not positioned at the center of the FOV,
this technique is insensitive to any excess.

\begin{table}
\caption{Supercuts gamma-ray selection criteria. These criteria were
optimized for each of the three camera configurations. Monte Carlo
simulations indicate that for observations at zenith angles $<
35^\circ$ this analysis results in an energy threshold of 350 GeV for
the $3.0^\circ$ and $3.5^\circ$ camera FOV and 500 GeV for the
$4.8^\circ$ camera FOV \cite{mohanty98}.}
\label{table:supercuts}
\vspace{0.5cm}
\begin{tabular}{ccc} \hline
Supercuts (1995/1996) & Supercuts (1997) & Supercuts (1998) \\
$3.0^\circ$ FOV       & $3.5^\circ$ FOV  & $4.8^\circ$ FOV  \\
\hline
\multicolumn{3}{c}{pre-selection criteria}                  \\ 
\hline
{\em max1} $>$ 100 d.c.$^a$		&
{\em max1} $>$ 95 d.c.			&
{\em max1} $>$ 78 d.c.  \\

{\em max2} $>$ 80 d.c.		        &	
{\em max2} $>$ 45 d.c.			&
{\em max2} $>$ 56 d.c.	\\

{\em size} $>$ 400 d.c. 	        &	
{\em size} $>$ 0 d.c.			&
{\em size} $>$ 0 d.c.	\\
\hline
\multicolumn{3}{c}{gamma-ray selection}                     \\
\hline
$0.073^\circ <$ {\em width}    $< 0.15^\circ$ &
$0.073^\circ <$ {\em width}    $< 0.16^\circ$ &
$0.073^\circ <$ {\em width}    $< 0.16^\circ$ \\

$0.16^\circ <$ {\em length}   $< 0.30^\circ$ &
$0.16^\circ <$ {\em length}   $< 0.33^\circ$ &
$0.16^\circ <$ {\em length}   $< 0.43^\circ$ \\

$0.51^\circ <$ {\em distance} $< 1.10^\circ$ &
$0.51^\circ <$ {\em distance} $< 1.17^\circ$ &
$0.51^\circ <$ {\em distance} $< 1.25^\circ$ \\

$\alpha < 15^\circ$		&	
$\alpha < 15^\circ$ 		&
$\alpha < 10^\circ$ \\

&
&
{\em asymmetry} $>0^\circ$ \\
\hline
\end{tabular}\\
a) d.c. = digital counts (1.0 d.c. $\approx$ 1.0 photoelectron).
\end{table}

\subsection{Photon Arrival Direction}
If the source position is ill-defined or if the source is extended,
the traditional method becomes ineffective. As a result two different
methods may be applied. In the first, the FOV is divided into a grid
and each grid point is tested for an excess of events above background
utilizing the Supercuts selection criteria given in
Table~\ref{table:supercuts} or using cuts which make use of {\em
asymmetry} to break the pointing ambiguity and {\em width/length} to
better localize the image. This method is described in detail in
\cite{akerlof91}, \cite{fomin94} and \cite{buckley98}. In the second
method, a unique arrival direction is determined thereby generating a
TeV gamma-ray map of the FOV (see also \cite{lebohec98} for a
description of an alternative method utilized by the Cherenkov Array at
Th\'emis (CAT) group). The benefit of the second method is that it can
be easily utilized in a search for extended or diffuse emission. This
second method is described below.

By employing several features of gamma-ray induced showers, the
arrival direction of a photon can be determined with the atmospheric
Cherenkov technique, using a single telescope. Cherenkov light images
of showers can be characterized as elongated ellipses with the major
axis representing the projection of the shower trajectory along the
image plane. The position of the source must lie along this axis near
the tip of the light distribution corresponding to the initial
interaction of the shower cascade. In addition, the position of the
source must lie in the direction indicated by the {\em asymmetry} of
the image. The elongation of a shower image and the angular distance
between its centroid and the source position depend upon the impact
parameter of the shower on the ground (see
Figure~\ref{figure:hillasparameters}). For small impact parameters,
the image should have a form close to that of a circle and be
positioned very near the source position in the focal plane
\cite{fomin94}. For increasing impact parameter it should become
elongated and have a form close to that of an ellipse and be
positioned further from the source position in the focal plane as
shown in Figure~\ref{figure:hillasparameters}. The elongation of an
image can be expressed as the ratio of its angular {\em width} and
{\em length}. A simple form for the relationship between the
elongation of an image and the angular distance between its centroid
and the source position, {\em disp}, is
\begin{equation}
	disp = \xi(1 - \frac{width}{length})
\label{equation:disp}
\end{equation}
where $\xi$ is a scaling parameter. A combination of these features
provides a unique arrival direction for each gamma-ray event.

\begin{figure}
\epsfig{file=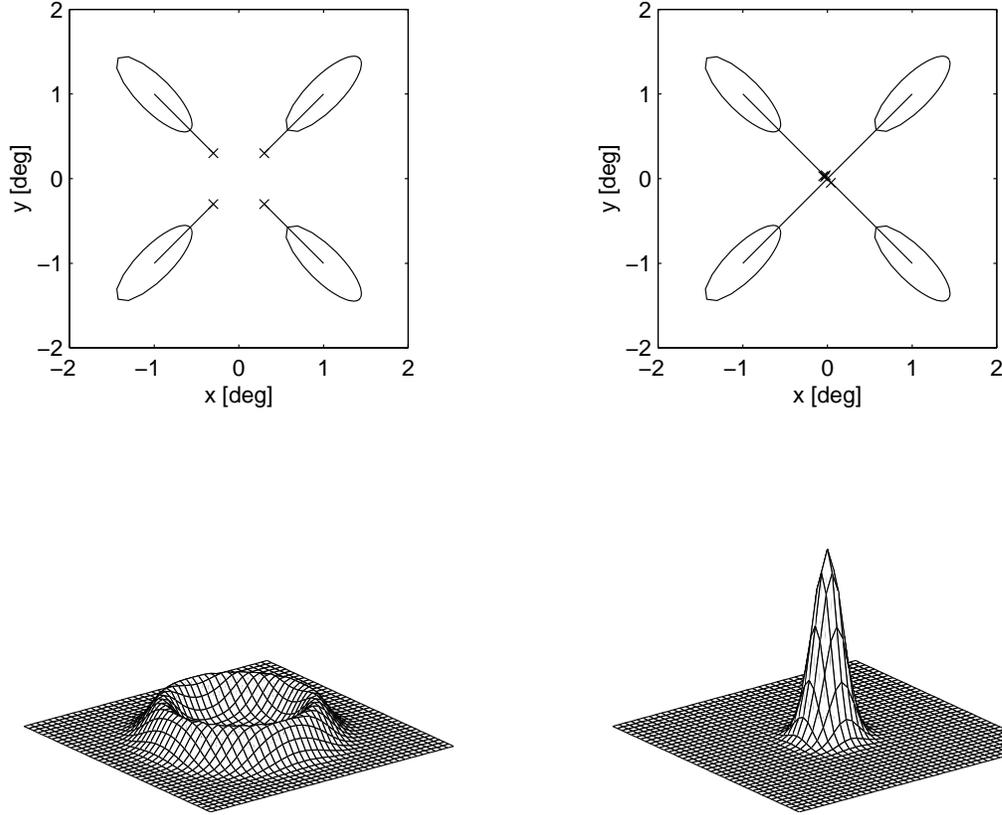,width=\linewidth}
\caption{Conceptual depiction of the angular spread of event arrival
directions. (Left) A value of $\xi$ which is too small so that the
arrival directions circle the source location. (Right) The optimum
value of $\xi$ minimizes the angular spread of arrival directions.}
\label{figure:ringoffire}
\end{figure}

The value of {\em disp} depends on an unknown scaling factor
$\xi$. Generally, $\xi$ will depend on the height of the shower in the
atmosphere, the elevation of the detector on the ground, the zenith
angle of the observation, parameters of the model of the atmosphere
and the energy of the primary particle. For observations at small
zenith angles ($< 35^\circ$), and for the majority of events near the
threshold of the instrument, the dependence on zenith angle and energy
can be neglected. The value of $\xi$ will be further modified by the
effect of the finite size of the camera in the focal plane giving rise
to edge effects due to the truncation of images. Experimentally, $\xi$
is determined from data so that the calculated shower arrival
directions line up with a known source position. For example, if a
value of $\xi$ is chosen which is too small, then due to azimuthal
symmetry, the arrival directions will circumvent the source location,
similarly for values of $\xi$ too large. The optimum value of $\xi$
minimizes the spread in the angular distribution of events centered at
the source position. The dependence of the angular spread of the
arrival directions on the scaling factor $\xi$ is depicted in
Figure~\ref{figure:ringoffire}. By minimizing the spread in the
arrival directions the optimum angular resolution is achieved. This
has the added advantage of increasing the signal to background ratio
when determining an excess of events from a region.

\begin{figure}
\epsfig{file=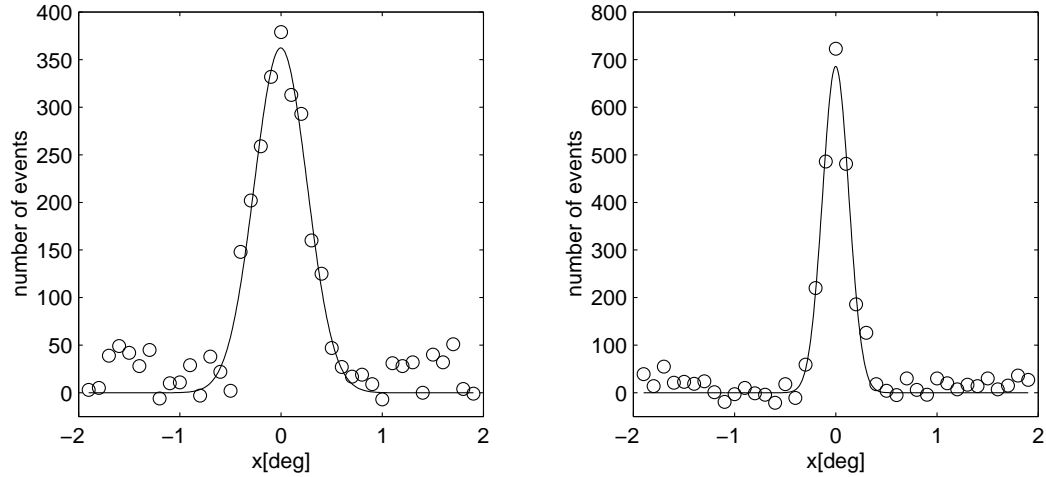,width=\linewidth}
\caption{Binned excess arrival directions projected along the x-axis
for scaling factor $\xi$ of 1.20 (left) and 1.78 (right).}
\label{figure:binarrival}
\end{figure}

Data from the Whipple 10 m gamma-ray telescope is used to optimize the
value of $\xi$. We include data from each of the three cameras (see
Table~\ref{table:dataset}), taken on established sources, positioned
at the center of the FOV which should appear point-like. First, events
are selected as gamma-like based on the angular {\em width} and {\em
length} criteria given in Table~\ref{table:supercuts}. No selection on
{\em distance} or $\alpha$ is made. Secondly, the arrival direction of
each event is determined utilizing Equation~\ref{equation:disp}. Next,
each arrival direction is corrected for the rotation of the FOV, due
to the altitude-azimuth mount used by the Whipple 10 m telescope, by
de-rotating the position to zero hour angle. These corrected arrival
directions are binned on a two dimensional grid of bin size
$0.1^\circ$x$0.1^\circ$. Figure~\ref{figure:binarrival} shows examples
of the binned excess arrival directions, for data taken on Markarian
501 (see Table~\ref{table:dataset}), projected along the x-axis, for
two values of the scaling parameter $\xi$. The excess shows the
difference between accumulated ON source events and corresponding OFF
source control data.

\begin{table}
\caption{Data used to optimize the scaling parameter $\xi$.}
\label{table:dataset}
\vspace{0.5cm}
\begin{tabular}{llll} \hline
Camera		& Source		& Observation	& Total ON			\\
FOV			& name		& period		& source time (mins)	\\
\hline
$3.0^\circ$		& Markarian 421
  \cite{punch92}  & 1995		& 523.2				\\
$3.5^\circ$		& Markarian 501	
  \cite{quinn96}  & 1997		& 471.6				\\
$4.8^\circ$		& Crab Nebula	
  \cite{weekes89} & 1998		& 845.9			\\
\hline
\end{tabular}
\end{table}		

\begin{figure}
\epsfig{file=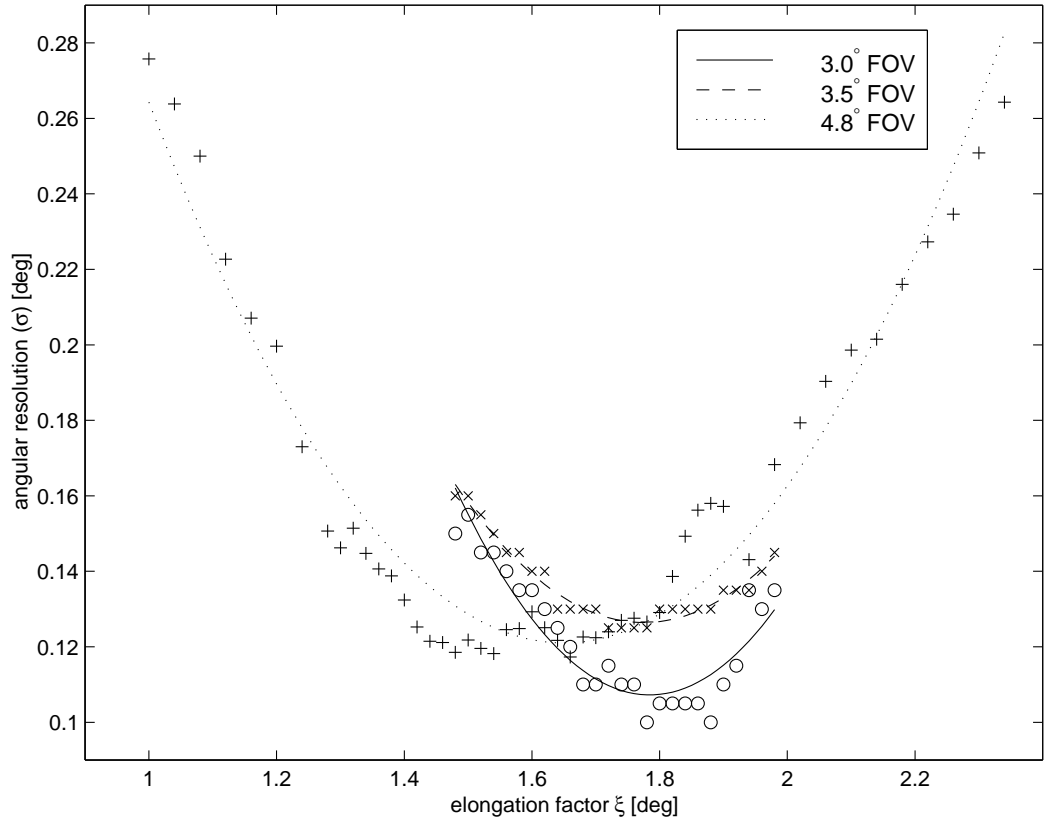,width=\linewidth}
\caption{Variation of the spread in excess arrival directions with
elongation scaling factor $\xi$. The data was fitted by a quadratic
function, shown by the curves, for the purpose of determining the
minimum spread.}
\label{figure:spreadarrival}
\end{figure}

The spread in arrival directions for each value of $\xi$ is determined
by fitting a Gaussian function of standard deviation $\sigma$ to the
binned excess. The results are depicted in
Figure~\ref{figure:spreadarrival}. The minimum spread occurs at a
value of $\xi = 1.78^\circ, 1.78^\circ, 1.65^\circ$ for the
$3.0^\circ, 3.5^\circ$ and $4.8^\circ$ camera FOV respectively. The
effect of the decreasing optimum value of $\xi$ with increasing FOV is
the result of less image truncation imposed by the larger camera. This
effect is also evident in the upper bound of the optimum length cut
used to select gamma-ray images (see Table~\ref{table:supercuts}). The
minimum spread is a measure of the optimum angular resolution, or
point spread function (PSF) of the technique and corresponds to
$\sigma = 0.11^\circ, 0.13^\circ$ and $0.12^\circ$ respectively.

\section{Results}
\begin{figure}
\epsfig{file=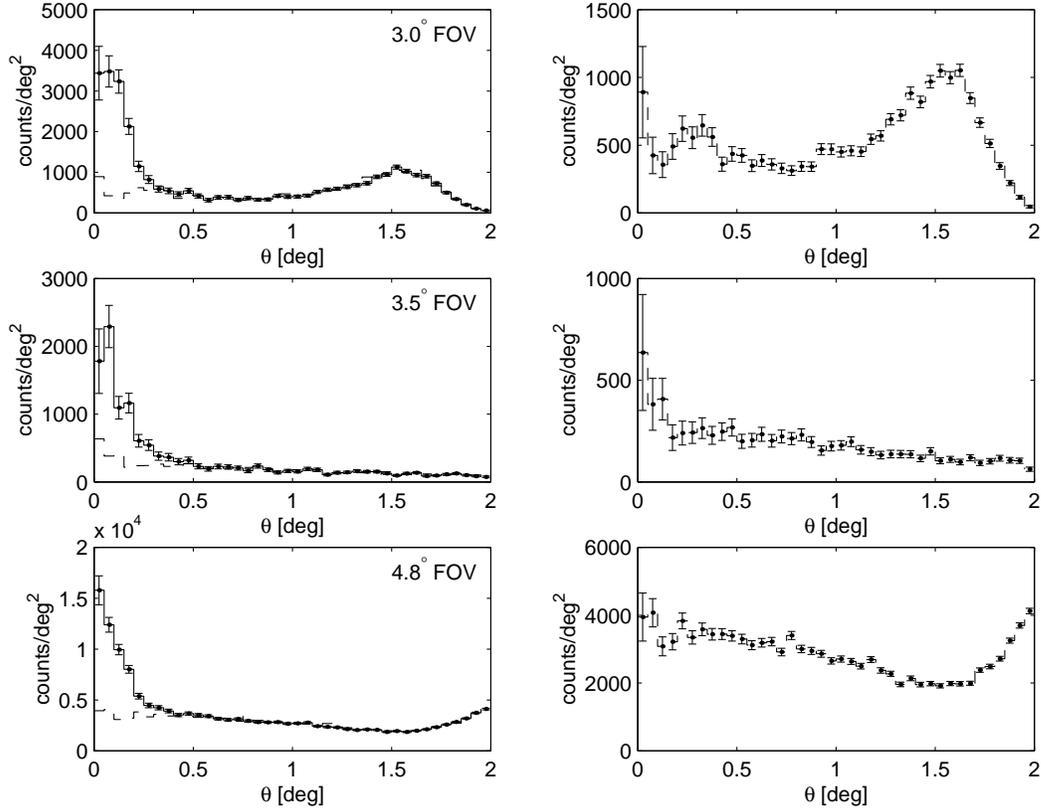,width=\linewidth}
\caption{Distribution of the radial distance of photon arrival
directions from the center of the FOV, $\theta$, for data taken with
all three cameras. (Left) The solid line depicts the ON source
observation and the dashed line shows the results of an equivalent
amount of time taken on a corresponding OFF source region. (Right) The
dashed line shows only the OFF source observation. The presence of a
rise in the number of events near the edge of the FOV is absent from
the $3.5^\circ$ camera results due to less image truncation imposed by
the smaller triggering FOV (see Figure~\ref{figure:camera}).}
\label{figure:radial}
\end{figure}

The determination of a photon's arrival direction enables the search
for emission from anywhere within the camera's FOV. When the source
position is known, a gamma-ray signal is detected as an excess of
events, between corresponding ON and OFF source observations,
originating from the source direction. Firstly, events are selected as
gamma-like based on the angular {\em width} and {\em length} criteria
given in Table~\ref{table:supercuts} and their arrival directions
determined as described above. Secondly, the radial distance from the
arrival direction to the source position, $\theta$, is calculated and
binned in $0.02^\circ$ intervals. The results of data taken with the
three cameras are shown in Figure~\ref{figure:radial}. The area of
exposure for each bin increases with greater radial distance,
therefore each bin contents have been divided by the area of the
annulus defined by the bin boundaries. The physical interpretation of
such a distribution is the surface brightness of an extended object.

\subsection{Aperture Selection Criteria}
\begin{figure}
\epsfig{file=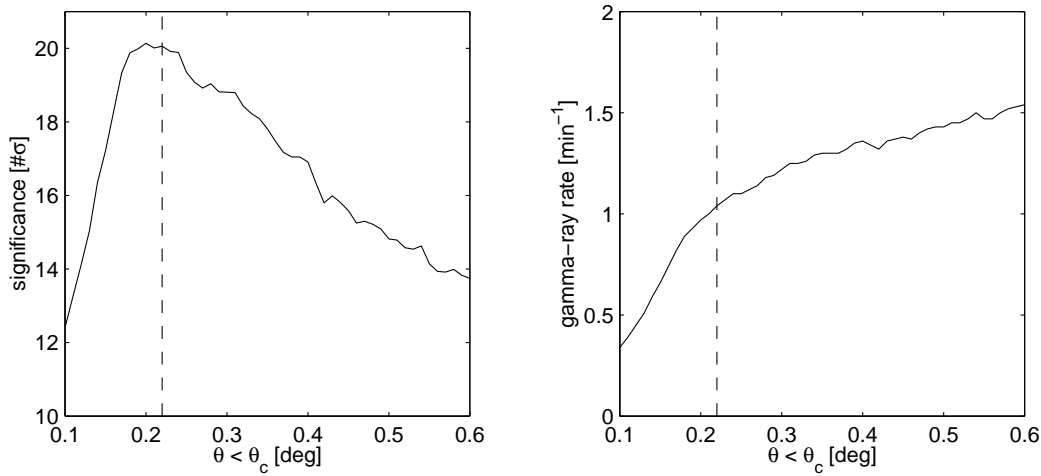,width=\linewidth}
\caption{Results of the optimization of the selection criteria $\theta
< \theta_c$. (Left) The solid line shows the statistical significance
of the excess between ON and corresponding OFF source
observations. The dashed line shows the value of the optimum cut
chosen, approximately the center of the peak. (Right) The solid line
shows the gamma-ray signal obtained.}
\label{figure:optrad}
\end{figure}

For each data set depicted in Figure~\ref{figure:radial}, an excess
gamma-ray signal is apparent in the ON source observation (solid line)
when compared to the OFF source observation (dotted line) and the
excess is confined to small values of $\theta$. The angular extent of
the signal is similar for each camera, consistent with the results of
the optimized angular resolution. The results of the OFF source
observations illustrate a background which appears uniform for small
values of $\theta$ with a gradual decline due to the finite size of
the camera. In the $3.0^\circ$ and $4.8^\circ$ cameras, all PMTs enter
into the trigger (all of the way out to the edge of the FOV), whereas
in the $3.5^\circ$ camera only the inner 91 out of the 151 PMTs enter
into the trigger (see Figure~\ref{figure:camera}).  Thus in the
$3.0^\circ$ and $4.8^\circ$ cameras there are a number of background
cosmic-ray events which lie on the edge of the camera and which are
distorted by image truncation. This truncation tends to produce images
whose major axes make an angle of $\alpha = 90^\circ$ and results in a
significant rise in the number of events collected at large values of
$\theta$. This rise is not apparent in the results from the
$3.5^\circ$ camera. The selection of events consistent with the
position of the source is accomplished by counting the number of
events within a circular aperture of radius $\theta_c$ centered at the
source position. The optimum value of $\theta_c$ is
determined using data taken with the $4.8^\circ$ camera. The results
are shown in Figure~\ref{figure:optrad} and yield an optimum selection
criteria of $\theta < 0.22^\circ$. This selection criteria can be
applied to the data for all camera configurations owing to the similar
radial extent of the gamma-ray signal. When the angular extent of the
source is of the order, or greater than the resolution of the
technique, the size of the circular aperture can be adjusted to the
size of the emission region. In doing so more background is included,
hence sensitivity is diminished. In background dominated counting
statistics, the sensitivity of a detector will be proportional to
$1/\sqrt{background}$. If we neglect the gradual decline in the
background with increasing angular offset, the sensitivity of an
atmospheric Cherenkov telescope utilizing this method will scale as
$1/\theta_c$.

\begin{figure}
\epsfig{file=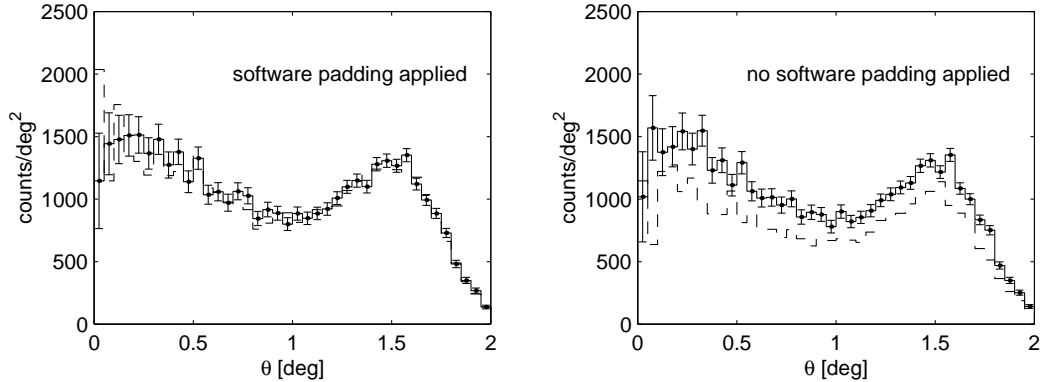,width=\linewidth}
\caption{Results of TeV gamma-ray observations of the diffuse supernova
remnant G78.2+2.1. Each panel depicts the distribution of $\theta$ for
the ON (solid line and error bars) and OFF (dashed line) source
observations.  The left panel depicts results utilizing software
padding. The right panel depicts results without the application of
software padding.}
\label{figure:gcygni_radial}
\end{figure}

TeV gamma-ray observations of the supernova remnant G78.2+2.1,
reported by the Whipple Collaboration in \cite{buckley98}, is an
example of an extended source. The expanding shell of the supernova
explosion subtends an angle of approximately $0.5^\circ$. The
distribution of $\theta$, for a subset of the data reported in
\cite{buckley98}, is given in Figure~\ref{figure:gcygni_radial}. In
the search for a gamma-ray signal, a circular aperture was chosen to
encompass the size of the emission region plus twice the angular
resolution to account for smearing of the edges.  As shown in
Figure~\ref{figure:gcygni_radial}, no significant excess was
obtained. Also shown is an example of a false signal resulting from
biases due to sky brightness differences. Without the application of
software padding, the brighter ON source region yields a greater
number of background events selected as gamma-rays resulting in a
significant excess between the ON and OFF source observations.

\subsection{Gamma-ray Collection Efficiency}
We compare the gamma-ray collection efficiency and background
rejection of the aperture selection method with the traditional
Supercuts criteria using data taken on the Crab Nebula, the standard
candle for TeV gamma-ray astronomy. The results from the three camera
configurations are given in Table~\ref{table:supercutsvsaperture}. The
data collected with the $3.0^\circ$ camera was taken in January -
February 1995. Like the traditional Supercuts criteria, the aperture
selection includes events which are oriented towards the center of the
FOV. However, the aperture selection has two further constraints in
that events must be skewed towards the center of the FOV and must be
elongated in proportion to the impact parameter on the ground. These
additional criteria result in greater background rejection compared
with Supercuts. Results from the $3.0^\circ$ camera indicate a reduced
gamma-ray collection efficiency due to image truncation imposed by the
smaller FOV. Truncated images reduce the effectiveness of the {\em
asymmetry} parameter by clipping the tails of the light distribution
which defines the skewness of the image. Data collected with the
$3.5^\circ$ and $4.8^\circ$ cameras were taken in January - February
1997 and November 1998 - January 1999 respectively.  First note that
for the data taken with the $4.8^\circ$ FOV camera, a degradation of
mirror reflectivity and lack of light cones resulted in an increased
energy threshold and lower count rate on the Crab Nebula.  Despite the
increased threshold this data can still be used to judge the
effectiveness of the extended camera. These results not only show the
same improvement in background rejection over the traditional
Supercuts criteria, but indicate similar gamma-ray collection
efficiency. This results in a marginal increase in sensitivity of the
aperture selection criteria over traditional Supercuts.

\begin{table}
\caption{Comparison of Supercuts with Aperture selection criteria
using data taken on the Crab Nebula (the standard candle for TeV
astronomy).}
\label{table:supercutsvsaperture}
\vspace{0.5cm}
\begin{tabular}{llllll} \hline
Selection&Observation &ON     &OFF    &Significance& Gamma-ray  \\
criteria &duration    &source &source &$\sigma$/hr & rate       \\
         &(min)       &counts &counts &            &(min$^{-1}$)\\
\hline
\multicolumn{6}{c}{$3.0^\circ$ FOV} \\
\hline
Supercuts& 246.9      &948    &408    & 7.2        & $2.2 \pm 0.1$  \\
Aperture &            &353    &77     & 6.6        & $1.12 \pm 0.08$\\ 
\hline
\multicolumn{6}{c}{$3.5^\circ$ FOV} \\
\hline
Supercuts& 304.6      &1654   &871    & 6.9        & $2.6 \pm 0.2$  \\
Aperture &            &790    &211    & 8.1        & $1.9 \pm 0.1$  \\
\hline
\multicolumn{6}{c}{$4.8^\circ$ FOV} \\
\hline
Supercuts& 845.9      &1683   &778    & 4.8        & $1.07 \pm 0.06$ \\
Aperture &            &1404   &524    & 5.3        & $1.04 \pm 0.05$ \\ 
\hline
\end{tabular}
\end{table}

\subsection{The Sky in TeV Gamma Rays}
\begin{figure}
\epsfig{file=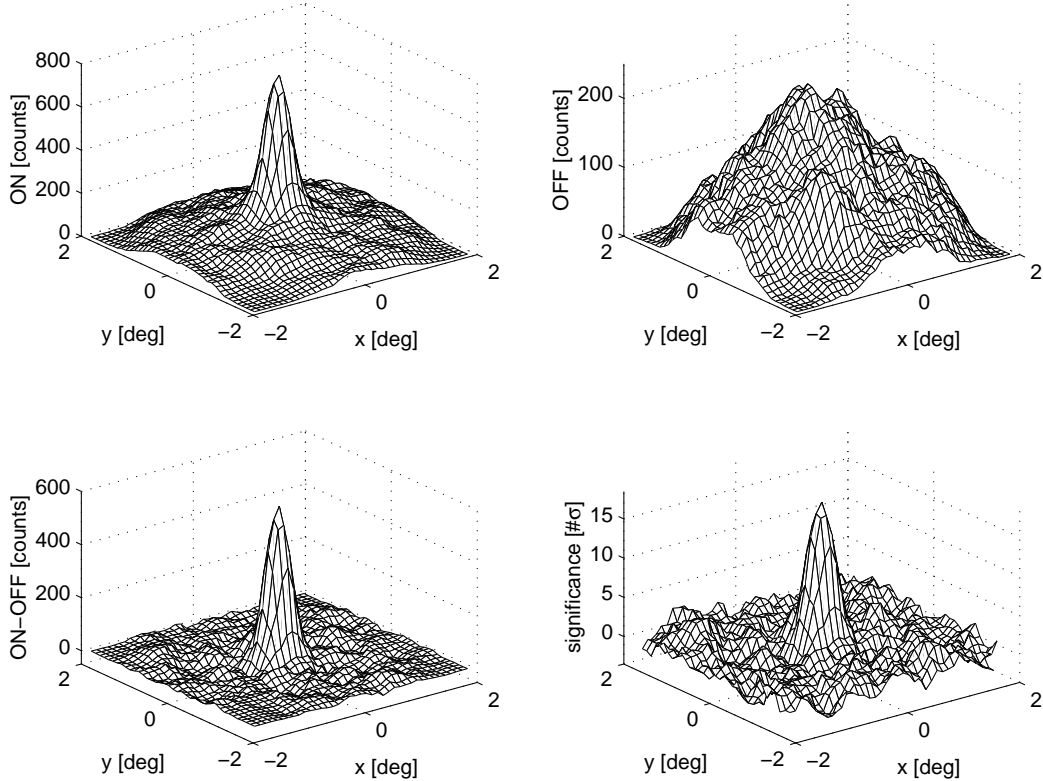,width=\linewidth}
\caption{Shown are the accumulated grid contents for the 
observations taken with the $3.5^\circ$ camera. (Top left) ON source
data given in Table~\ref{table:dataset}. (Top right) The corresponding
OFF source data. (Bottom left) The difference in counts and
significance (Bottom right).}
\label{figure:grid}
\end{figure}

If the position of the source is unknown but assumed to be in the FOV,
a two dimensional grid of bin size $0.1^\circ$x$0.1^\circ$ is
constructed.  Events are first selected as gamma-like based on the
angular {\em width} and {\em length} criteria given in
Table~\ref{table:supercuts}, and their arrival directions determined
as described above. Next, the contents of each grid point within the
optimum cut on the radial distance from the arrival direction is
incremented by one. Figure~\ref{figure:grid} shows the accumulated
grid contents for data taken with the $3.5^\circ$
camera. Conceptually, this performs a type of image smoothing about
the calculated photon arrival direction resulting in an oversampled
gamma-ray map which is tuned to the search for point sources anywhere
within the camera's FOV. The statistical significance of excess events
between ON and corresponding OFF source observations is calculated for
each grid point using the method of \cite{li83}. The use of Poissonian
statistics is valid in this case because grid points are incremented
by unity. However, one must be careful in interpreting the
significance.  To be precise, the test statistic calculated at each
grid point can only be interpreted as significance if there was a
prior hypothesis for gamma-ray emission from a point source at that
position. Otherwise, the number of independent grid points tested, or
trials factor, must be taken into account thereby degrading the
significance. For example, for a grid ($3.9^\circ$ FOV divided into
$0.1^\circ$ bins) with 1521 grid points (not independent) we have
effectively 324 trials and expect on average one $3\sigma$ excess.

\begin{figure}
\epsfig{file=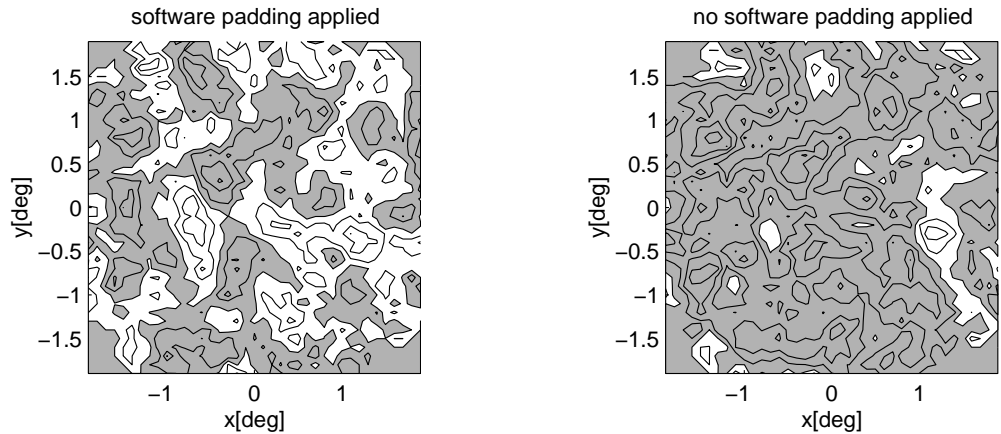,width=\linewidth}
\caption{Results of TeV gamma-ray observations of the supernova remnant
G78.2+2.1. The contours are proportional to the statistical
significance of the smoothed excess gamma-ray events and increment by
one standard deviation. Contours depicting positive excess are shaded
in gray.  The left panel depicts results utilizing software
padding. The right panel depicts results without the application of
software padding.}
\label{figure:gcygni_image}
\end{figure}

An example of a TeV gamma-ray image of a region of sky (the supernova
remnant G78.2+2.1), found to be devoid of a gamma-ray signal, is shown
in Figure~\ref{figure:gcygni_image}.  The image shows several
$2\sigma$ excesses and one $3\sigma$ excess.  However, without a prior
hypothesis for gamma-ray emission from a point source at these
positions a trials factor for the number of independent grid points
must applied, resulting in a reduction in significance.  Also shown in
Figure~\ref{figure:gcygni_image} is an example of a false signal
resulting from biases due to sky brightness differences. The region of
the image showing the greatest excesses appear well correlated with
the sky brightness difference between the ON and OFF source regions as
depicted in Figure~\ref{figure:gcygni_pedvar}.

\begin{figure}
\epsfig{file=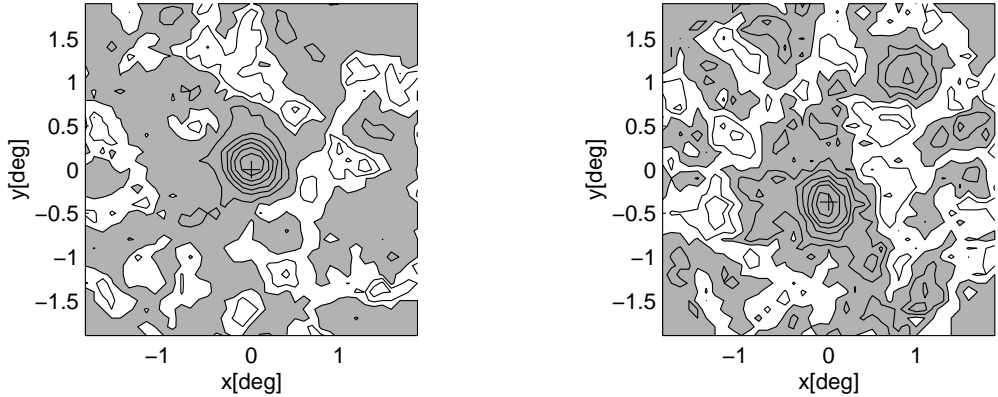,width=\linewidth}
\caption{Results of the observations taken with $3.0^\circ$ camera 
on the Crab Nebula, centered in the FOV (Left) and offset in
declination by 22 arcminutes (Right), corresponding to observation
times of 247 min and 83 min respectively. The contours are proportional to
the statistical significance of the smoothed excess gamma-ray events
and increment by two standard deviations (Left) and one standard
deviation (Right). Contours depicting positive excess are shaded
in gray. The cross indicates the position of the Crab Nebula
and the systematic pointing uncertainty.}
\label{figure:craboffset3.0}
\end{figure}

\begin{figure}
\epsfig{file=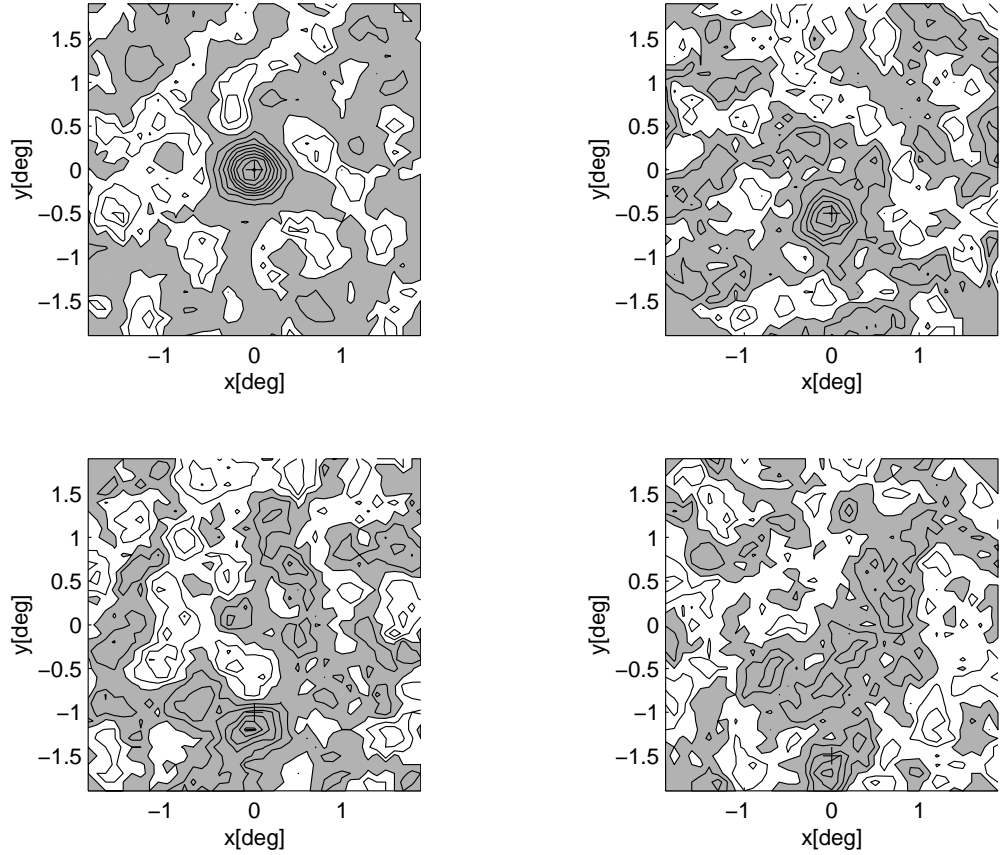,width=\linewidth}
\caption{Results of the observations taken with $3.5^\circ$ camera 
on the Crab Nebula, centered in the FOV (Top Left) and offset in
declination by $0.5^\circ$ (Top Right), $1.0^\circ$ (Bottom Left) and
$1.5^\circ$ (Bottom Right), corresponding to observation times of 305
min, 55 min, 83 min and 138 min respectively. The contours are
proportional to the statistical significance of the smoothed excess
gamma-ray events and increment by two standard deviations for the
centered observation and by one standard deviation for the offset
observations. Contours depicting positive excess are shaded in
gray. The cross indicates the position of the Crab Nebula and the
systematic pointing uncertainty.}
\label{figure:craboffset3.5}
\end{figure}

\begin{figure}
\epsfig{file=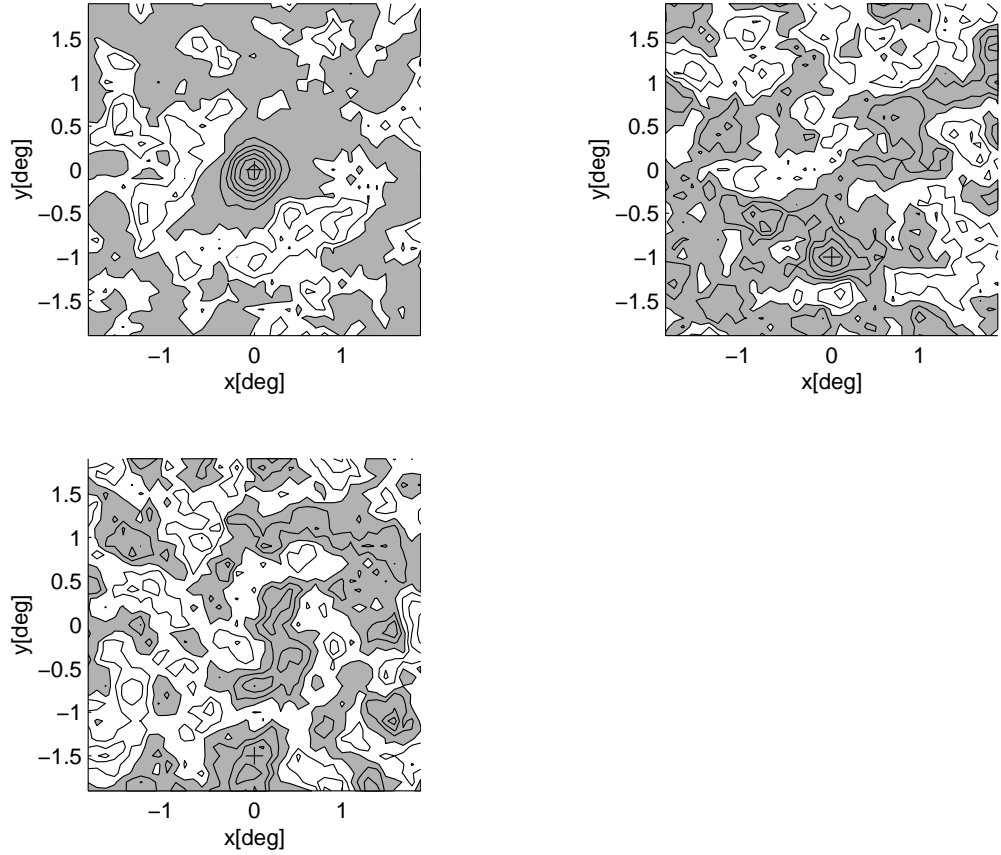,width=\linewidth}
\caption{Results of the observations taken with $4.8^\circ$ camera 
on the Crab Nebula, centered in the FOV (Top Left) and offset in
declination by $1.0^\circ$ (Top Right) and $1.5^\circ$ (Bottom Left),
corresponding to observation times of 846 min, 55 min and 107 min
respectively. The contours are proportional to the statistical
significance of the smoothed excess gamma-ray events and increment by
three standard deviations for the centered observation and by one
standard deviation for the offset observations. Contours depicting
positive excess are shaded in gray. The cross indicates the position
of the Crab Nebula and the systematic pointing uncertainty.}
\label{figure:craboffset4.8}
\end{figure}

To test the efficacy of the technique and measure the efficiency and
sensitivity away from the center of the FOV, we analyzed data taken on
the Crab Nebula which was offset in declination. The results are shown
in Figures~\ref{figure:craboffset3.0},\ref{figure:craboffset3.5} and
\ref{figure:craboffset4.8} for the $3.0^\circ, 3.5^\circ$ and
$4.8^\circ$ cameras respectively. The contours are proportional to the
statistical significance of the excess between the ON and OFF source
data. The peak excess and detected gamma-ray rates are given in
Table~\ref{table:offsetresults}.

\begin{table}
\caption{Results of data taken on the Crab Nebula offset in declination.}
\label{table:offsetresults}
\vspace{0.5cm}
\begin{tabular}{lllll} \hline
Offset     & Observation    & Standard deviation & Significance & Gamma-ray\\
($^\circ$) & duration (min) & of PSF ($^\circ$)  & \#$\sigma$   & rate (min$^{-1}$) \\ 
\hline
\multicolumn{4}{c}{$3.0^\circ$ FOV} \\
\hline
0.00       & 246.9          & 0.15         & 13.4         & $1.08 \pm 0.08$ \\
0.37       & 83.0           & 0.13         & 6.8          & $0.9 \pm 0.1$   \\
\hline
\multicolumn{4}{c}{$3.5^\circ$ FOV} \\
\hline
0.00       & 304.6          & 0.13         & 18.3         & $1.9 \pm 0.1$   \\
0.50       & 54.8           & 0.13         & 6.1          & $1.2 \pm 0.2$   \\
1.00       & 83.2           & 0.15         & 6.3          & $0.8 \pm 0.1$   \\
1.50       & 138.5          & 0.11         & 4.3          & $0.35 \pm 0.08$ \\
\hline
\multicolumn{4}{c}{$4.8^\circ$ FOV} \\
\hline
0.00       & 845.9          & 0.12         & 20.1         & $1.04 \pm 0.05$ \\
1.00       & 55.4           & 0.10         & 4.5          & $0.7 \pm  0.2$  \\
1.50       & 106.7          & 0.14         & 3.7          & $0.35 \pm 0.09$ \\
\hline
\end{tabular}
\end{table}

\begin{figure}
\epsfig{file=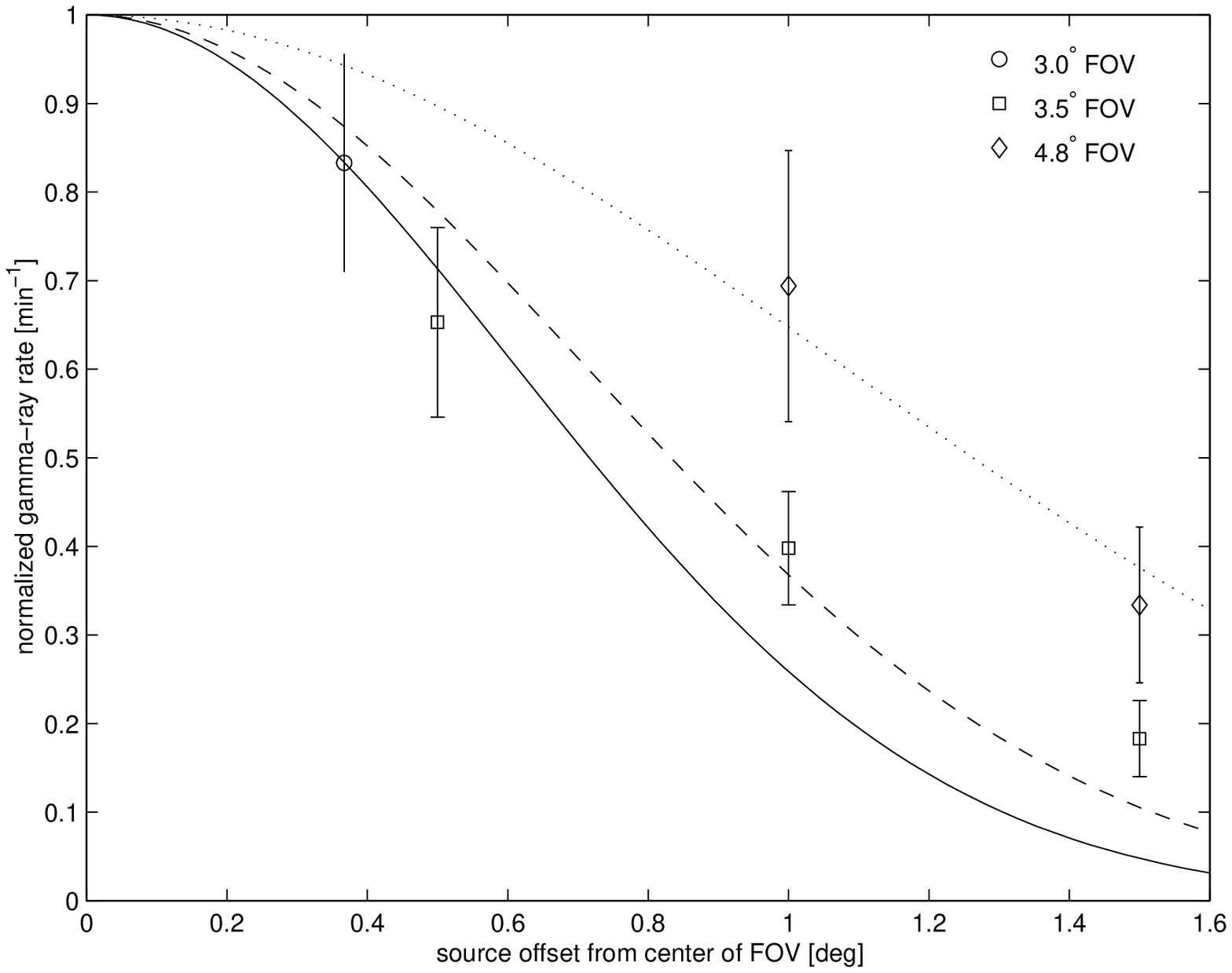,width=\linewidth}
\caption{The measured gamma-ray rate from the Crab Nebula normalized
to the rate at the center of the FOV. The curves indicate the general
trend of the data.}
\label{figure:craboffset}
\end{figure}

The results show that the technique is sensitive to a gamma-ray signal
offset from the center of the FOV. For the $3.5^\circ$ and $4.8^\circ$
cameras, the Crab Nebula was detected with a sensitivity of
approximately $3\sigma$/hr for an offset of $1.5^\circ$. The derived
position of the Crab Nebula was within the $0.1^\circ$ pointing
accuracy of the Whipple 10 m telescope for offsets less than
$1.0^\circ$. We note that there appears to be a small systematic shift
of the derived position of the Crab Nebula for the $1.0^\circ$ and
$1.5^\circ$ offsets for the data taken with the $3.5^\circ$ camera and
for the $1.5^\circ$ offsets for the data taken with the $4.8^\circ$
camera. The shift is most likely due to the inclusion of truncated
images as the source moves closer to the edge of the FOV.  The spread
of events about the derived source position was determined by fitting
a Gaussian function to the unsmoothed binned excess. The results
indicate that the PSF is approximately constant over the FOV of the
camera and thus not dependent on the optical properties of the
reflector which degrade with increasing angular offset.

As the gamma-ray source moves away from the center of the camera, less
of the Cherenkov light pool falls on the detector at a given impact
parameter. This has the effect of reducing the collection area for
gamma-rays. The measurements of the gamma-ray rate from the Crab
Nebula at increasing angular offset is a direct measure of this
reduction as shown in Figure~\ref{figure:craboffset}. The effective
FOV, which we arbitrarily define as the radius corresponding to a 50\%
efficiency, appears to scale linearly with camera physical FOV, at
least up to the range of offsets and camera sizes included in these
results. For example, a camera with a physical FOV of $6.0^\circ$
would have an effective FOV double that of a camera with a $3.0^\circ$
FOV. Moreover, by doubling the diameter of the FOV a fourfold increase
in sensitive area for sources offset from the center of the FOV
is realized.

\section{Conclusion}
We have described a method of atmospheric Cherenkov imaging which
reconstructs the unique arrival direction of TeV gamma-rays using a
single telescope. This method is derived empirically making use of the
Crab Nebula as a standard candle to optimize the angular resolution of
the technique.  This allows a selection of events based on the
position of a source anywhere within the telescope's FOV. We found
that such a selection yields similar sensitivity and collection
efficiency to the traditional Supercuts criteria utilized by the
Whipple Collaboration. However, as demonstrated, the technique is
easily applicable to sources offset from the center of the FOV or
sources of extended emission.

\ack{This research is supported by grants from the U.S. Department of
Energy. We are grateful to Dave Fegan and Trevor Weekes for their
guidance and assistance. We thank the Whipple Gamma-ray Collaboration
for the use of the data presented in this paper and acknowledge the
technical assistance of K. Harris and E. Roache.}


\begin{thebibliography}{999}
\bibitem{weekes72} T.C. Weekes, et al., {\em ApJ}, {\bf 174} (1972) 165.
\bibitem{cawley90} M.F. Cawley, et al., {\em Exper. Astr.}, {\bf 1} (1990) 173.
\bibitem{hillas85} A.M. Hillas, Proc. 19th ICRC, {\bf 3} 
(La Jolla, USA, 1985) 445.
\bibitem{fegan94} D.J. Fegan, et al., Proc. Towards a Major Atmospheric 
Cherenkov Detector III (Tokyo, Japan, 1994) 149.
\bibitem{connaughton98} V. Connaughton, et al., {\em APh}, {\bf 8} (1998) 179.
\bibitem{buckley98} J.H. Buckley, et al., {\em A\&A}, {\bf 329} (1998) 639.
\bibitem{lebohec2000} S. Le Bohec, et al., {\em ApJ}, in press (2000).
\bibitem{cawley93} M.F. Cawley, Proc. Towards a Major Atmospheric 
Cherenkov Detector II (Calgary, Canada, 1993) 176.
\bibitem{reynolds93} P.T. Reynolds, et al., {\em ApJ}, {\bf 404} (1993) 206.
\bibitem{catanese95} M. Catanese, et al., Proc. Towards a Major Atmospheric 
Cherenkov Detector IV (Padova, Italy, 1995) 335.
\bibitem{mohanty98} G. Mohanty, et al., {\em APh}, {\bf 9} (1998) 15.
\bibitem{akerlof91} C.W. Akerlof, et al., {\em ApJ}, {\bf 377} (1991) L97.
\bibitem{fomin94} V.P. Fomin, et al., {\em APh}, {\bf 2} (1994) 137.
\bibitem{lebohec98} S. Le Bohec, et al., {\em Nucl. Instr. and Meth. A}, 
{\bf 416} (1998)
425.
\bibitem{punch92} M. Punch, et al., {\em Nature}, {\bf 358} (1992) 477.
\bibitem{quinn96} J. Quinn, et al., {\em ApJ}, {\bf 456} (1996) L83.
\bibitem{weekes89} T.C. Weekes, et al., {\em ApJ}, {\bf 342} (1989) 379.
\bibitem{li83} T.P. Li and Y.Q Ma, {\em ApJ} {\bf 272} 317.
\end{thebibliography}
\end{document}